\documentclass[useAMS,usenatbib,usegraphicx]{mn2e}
\usepackage{hyperref}
\usepackage{color}

\title[Direct detection of Black Holes]{Direct detection of Black Holes via electromagnetic radiation}
\author[J. L. G. Sobrinho and P. Augusto]{J. L. G. Sobrinho$^{1,2}$\thanks{E-mail: sobrinho@uma.pt (JLGS)} and P. Augusto$^{1, 2}$\thanks{E-mail: sciman@med.up.pt (PA)}\\
$^{1}$Centro de Ci\^{e}ncias Exactas e da Engenharia, Universidade da Madeira, Caminho da Penteada, 9000-390 Funchal, Portugal\\
$^{2}$Centro de Astronomia e Astrof\'{i}sica da Universidade de Lisboa, Tapada da Ajuda, Edif\'{i}cio Leste, 2$^{\circ}$ piso, 1349-018 Lisboa, Portugal}

\begin{document}

\date{Accepted 2014 April 21. Received 2014 April 21; in original form 2013 April 23}

\pagerange{\pageref{firstpage}--\pageref{lastpage}} \pubyear{2014}

\maketitle

\label{firstpage}

\begin{abstract}
Many black hole (BH) candidates exist, ranging from supermassive ($\sim10^{6}$--$10^{10}$~M$_{\sun}$) to stellar masses ($\sim 1$--$100$~M$_{\sun}$), all of them identified by indirect processes. Although there are no known candidate BHs with sub-stellar masses, these might have been produced in the primordial Universe. 
BHs emit radiation composed of photons, gravitons and, later in their lifes, massive particles. 
 
We explored the detection of such BHs with  present day masses from $10^{-22}$~M$_{\sun}$ to $10^{-11}$~M$_{\sun}$. We determined the maximum distances ($d$) at which the current best detectors should be placed in order to identify such isolated BHs. Broadly, we conclude that in the visible and ultraviolet BHs can be directly detected at $d\la 10^7$~m while in the X-ray band the distances might reach $\sim10^8$~m (of the order of the Earth-Moon distance) and in the $\gamma$-ray band BHs might even be detected from as far as $\sim 0.1$~pc. 

Since these results give us realistic hopes of directly detecting  BHs, we suggest the scrutiny of current and future space mission data to reach this goal.
\end{abstract}

\begin{keywords}black hole physics.\end{keywords}

\section{Introduction}
\label{sec:Introduction}

Black Holes (BHs) are objects predicted by the laws of Physics: they naturally arise by solving the field equations of General Relativity. In terms of mass, a BH can be classified as supermassive ($\sim10^{6}$--$10^{10}$~M$_{\sun}$), intermediate ($\sim10^{3}$--$10^{5}$~M$_{\sun}$), stellar ($\sim 1$--$100$~M$_{\sun}$) or sub-stellar ($<1$~M$_{\sun}$). It is now well-established that supermassive BHs (SMBHs) reside in the centres of galaxies \citep[e.g.][]{2009MNRAS.393..838N} including our own galaxy with a $4.4\times10^{6}$~M$_{\sun}$ SMBH \citep*[e.g.][]{2010RvMP...82.3121G}. Intermediate mass BHs might form either in the core of star clusters \citep[e.g.][]{2004Natur.428..724P} or  galaxies \citep[e.g.][]{2004ApJ...610..722G}. These BHs can be detected indirectly by studying the gas and star dynamics on their surroundigs. In fact, a few have now been found \citep[e.g.][]{Greene2012}.

A stellar mass BH is the likely remnant of an exploded star with an initial mass greater than $\sim40$--$60$~M$_{\sun}$. It is best detected if it is part of a binary system with a giant star as companion from studying its dynamics. Another path to detection arises when the giant star gas, if fueling a BH, forms an accretion disk, the inner parts of which shine brightly in X-rays and $\gamma$-rays. With an estimated mass of $14$--$16$~M$_{\sun}$, the X-ray source Cyg X1 is the strongest candidate for a stellar mass BH in a binary system \citep[][]{2011ApJ...742...84O}.

It seems that the only way to form BHs with masses smaller than about 3~M$_{\sun}$ in the present Universe is in accelerators such as the \emph{Large Hadron Collider} \citep*[e.g.][]{2001PhRvL..87p1602D,2003CQGra..20L.205C} or, eventually, when cosmic-rays collide with the upper layers of the atmosphere \citep[e.g.][]{2002PhRvD..65l4027A}. However, in both cases, the detection of BHs (with a few Planck masses -- $M_P\sim 10^{-8}\mathrm{~kg}$) might only take place within the framework of extra dimensions. 
Rather more promising is the possibility of these smaller BHs ($<3$~M$_{\sun}$) having been formed during the early stages of the Universe (in which case they are called Primordial Black Holes (PBHs)) due, for example, to the gravitational collapse of density fluctuations \citep[][]{1971MNRAS.152...75H}.

\citet{1974Natur.248...30H} has shown that, when quantum effects are taken into account, BHs radiate. This process hardly affects stellar mass BHs, but it could be very significant in the case of smaller mass PBHs. In fact, PBHs with initial masses $\sim 10^{15}\mathrm{g}$ ($\sim 10^{-18}$~M$_{\sun}$) might  be exploding right now \citep[e.g.][]{1976ApJ...206....1P, 1976ApJ...206....8C, 1991ApJ...371..447M, 2000APh....12..269B, 2010PhRvD..81j4019C} while less massive ones should already have completely evaporated. PBHs with initial masses $\ga 10^{15}\mathrm{g}$ ($\ga 10^{-18}$~M$_{\sun}$) might  still be lurking around us, evaporating and/or accreting matter \citep[][]{sobrinho2011}.

In this paper we explore the pathway for direct BH detection which, if someday successful, could revolutionize the way we identify BHs in the Universe. We finalize by suggesting the use of space mission data in order to achieve this.

\section{Black Hole Thermodynamics}
\label{sec:Black Hole Thermodynamics}

\subsection{Primary emission}

A BH is completely characterized by only three parameters: mass $m$, electric charge $\epsilon$ and angular momentum per unit mass $a$. These parameters must follow the relation $a^{2}+\epsilon^{2}\leq m^{2}$ \citep[e.g.][]{1978RPPh...41.1313D} with $m$, $\epsilon$, and $a$ written in geometrized units ($c=G=1$; $G$ is the gravitational constant; $c$ the speed of light). There is a remarkable analogy between the laws of BH mechanics and the laws of Thermodynamics \citep[e.g.][]{wal98}. Thus, it is possible to assign to each BH a temperature $T=\hbar \Psi/(2\pi k)$ where $\hbar$ is the reduced Planck constant, $k$ is the Boltzmann constant and
\begin{equation}
\label{equ-gravsup}
\Psi=\frac{\left(m^{2}-a^{2}-\epsilon^{2}\right)^{1/2}}
{2m\left(m+\left(m^{2}-a^{2}-\epsilon^{2}\right)^{1/2}\right)-\epsilon^{2}}, \:\: (\epsilon\neq0, a\neq0)
\end{equation}
is the surface gravity of the BH \citep[e.g.][]{wal84}. For a given mass $m$, we  have $\Psi \propto T$, which will be maximum if $\epsilon$ and $a$ are both zero, i.e., if we have a Schwarzschild BH ($\Psi_{max}=\frac{1}{4m}$, $T_{max}=\frac{\hbar}{8\pi km}$). Reissner--Nordstr\"{o}m BHs ($m\neq 0,\epsilon \neq 0,a=0$), Kerr BHs ($m\neq 0,\epsilon=0 ,a\neq 0$) or Kerr--Newmann BHs ($m\neq 0,\epsilon \neq 0 ,a\neq 0$) with the same mass $m$ will have lower temperatures. In fact, for an extreme Reissner--Nordstr\"{o}m BH ($\epsilon=\pm m$) or a maximum Kerr BH ($a=\pm m$) the temperature is zero. Thus, for a given mass $m$, Schwarzschild BHs are the hottest ones and, consequently, the ones that offer more hypotheses of detection. We shall consider then, from now on, only Schwarzschild BHs.

The Schwarzschild BH has a central singularity and an infinite redshift surface (event horizon) with a dimension interpreted as radius (Schwarzschild radius) given by $r_{s}=2m=2GM/c^{2}$ where $M$ is the BH mass in non-geometrized units. The Schwarzschild BH temperature can be written, in non-geometrized units, as \citep[e.g.][]{1985reas.book.....D}
\begin{equation}
\label{equ-temp2}
T=\frac{\hbar c^{3}}{8\pi kGM}\approx 6.2\times10^{-8}\frac{M_{\sun}}{M} \:\: [\mathrm{K}] \,.
\end{equation}
We may relate this temperature with the wavelength $\lambda_{max}$ at which the emission gets its intensity peak since we have, according to Wien's Displacement Law \citep[e.g.][]{1985qpam.book.....E}:
\begin{equation}
\label{equ-wien}
T\lambda_{max}=2.898\times10^{-3} \: \: [\mathrm{Km}] \,.
\end{equation}
A Schwarzschild BH gets completely defined given its mass $M$: all other properties (e.g. radius, temperature) can be expressed in terms of $M$. In particular, knowing $M$, we know the value of $\lambda_{max}$ (cf. equations (\ref{equ-temp2}) and (\ref{equ-wien})) which means that we can associate to each wavelength of the electromagnetic spectrum a Schwarzschild BH which emits more intensely there. In this sense, {\em BHs have colours}. Hence, depending on where their electromagnetic emission spectrum peak is located, we define radio, infrared, visible, ultraviolet, X-ray and $\gamma$-ray BHs.

A Schwarzschild BH emits particles with energy in the range $(E,E+dE)$ at a rate \citep[e.g.][]{1991Natur.353..807H}
\begin{equation}
\label{particle_emission}
\frac{d^{2}N}{dEdt}=\frac{\Gamma_{s}}{2\pi \hbar }\left[\exp\left(\frac{8\pi GME}{\hbar c^3}\right)-(-1)^{2s}\right]^{-1}
\end{equation}
per degree of particle freedom. Here $s$ is the particle spin, $M$ the mass of the BH and $\Gamma_{s}$ the dimensionless absorption coefficient which is, in general, a function of $s$, $M$ and $E$. 
In the low-energy limit ($GME/\hbar c^3\ll 1$) we have \citep[][]{1990PhRvD..41.3052M}
\begin{eqnarray}
\label{GAMMA-spin1}
\Gamma_{s}(M,E)_{E\rightarrow 0}\approx \frac{64G^{4}M^{4}E^{4}}{3\hbar ^{4}c^{12}} & (s=1)
\end{eqnarray}
\begin{eqnarray}
\label{GAMMA-spin1half}
\Gamma_{s}(M,E)_{E\rightarrow 0}\approx \frac{2G^{2}M^{2}E^{2}}{\hbar ^{2}c^{6}} & (s=1/2)
\end{eqnarray}
At high energies the spectra approaches the blackbody emission and we get for all kind of particles \citep[][]{1990PhRvD..41.3052M}
\begin{equation}
\label{GAMMA-high}
\Gamma_{s}(M,E)_{E\rightarrow \infty}\approx \frac{27G^{2}M^{2}E^{2}}{\hbar ^{2}c^{6}}
\end{equation}
In the case of the photon ($s=1$) equation (\ref{particle_emission}) becomes
\begin{equation}
\label{particle_emission_photon0}
\frac{d^{2}N}{dEdt}=\frac{\Gamma_{s}}{2\pi\hbar}\left(\exp\left(\frac{8\pi GME}{\hbar c^3}\right)-1\right)^{-1}
\end{equation}
which gives us the number of photons emitted per unit energy, per unit time. Considering that $E=h\nu$ this becomes
\begin{equation}
\label{particle_emission_photon}
\frac{d^{2}N}{d\nu dt}=\Gamma_{s}\left(\exp\left(\frac{8\pi GME}{\hbar c^3}\right)-1\right)^{-1}
\end{equation}
which gives us the number of photons emitted per unit frequency, per unit time.

\subsection{$\gamma$-ray emission}

As a BH radiates due to the Hawking process it looses mass  or, in other words, it evaporates. The rate at which a BH evaporates can be written as \citep*[e.g.][]{1996PhRvL..76.3474M}
\begin{equation}
\label{equ-masslossrate}
\frac{dM}{dt}=-\frac{5.34\times10^{16}f(M)}{M^{2}} \: \: [\mathrm{kgs}^{-1}] \,
\end{equation}
where $f$ is a non-dimensional function of the BH mass $M$ accounting for the contributions of the different species of particles being emitted by the BH \citep[e.g.][]{1991PhRvD..44..376M,1997IJMPA..12.4167G}. When $M\gg 10^{14}$~kg we have $f(M)\approx 1$ \citep[e.g.][]{1996PhRvL..76.3474M} and for $M\ll 10^{8}$~kg we have $f(M)\approx 15.4$ \citep[e.g.][]{1994ApJ...436..254S}. Taking into account that $f(M)$ varies very slowly with $M$ \citep[e.g.][]{2002ApJ...568L...1H} we can integrate equation (\ref{equ-masslossrate}) in order to obtain
\begin{equation}
\label{equ-evaptime}
t_{evap}\approx \frac{M_{i}^{3}-M_{f}^{3}}{1.6\times10^{17}f(M)} \: \: [\mathrm{s}] 
\end{equation}
which is the time required for a BH to change its mass from the initial value $M_{i}$ to the  final value  $M_{f}$ ($M_f < M_i$). 
 
As the evaporation goes on, the BH will start emitting massive particles, besides photons and gravitons  \citep[][]{1976ApJ...206....1P}. In particular, an evaporating BH starts emitting hadrons, beginning with the lightest ones which are the $\pi^{0}$ mesons ($m_0\simeq 2.4\times10^{-28}$~kg) when its mass is $\sim 10^{12}$~kg  which corresponds to a Schwarzswchild radius of the order of the strong nuclear force range ($\sim 10^{-15}$~m). Because of that, the BH will emit jets of quarks and gluons instead of composed particles \citep[][]{1976ApJ...206....1P}. Emitted quarks and gluons develop into hadron jets with a predominance of $\pi$ mesons \citep[e.g.][]{1994ApJ...436..254S}. 

All three known kinds of $\pi$ mesons appear in a jet with the same probability. Every $\pi^{+}$ or $\pi^{-}$ meson decays into electrons, positrons and neutrinos. As for the $\pi^{0}$ mesons, each one of them decays into two $\gamma$-ray photons \citep[e.g.][]{1994ApJ...436..254S}.
We call these {\em secondary $\gamma$-rays} in contrast to the $\gamma$-rays emitted directly by the BH which we call {\em primary $\gamma$-rays}.

In order to properly determine the global $\gamma$-ray spectrum emitted by the BH, equation (\ref{particle_emission}) must be convolved with the fragmentation function for pions which gives the number of pions produced in the energy interval $(E,E+dE)$ due to a quark-gluon jet of energy Q \cite[][]{1990PhRvD..41.3052M}. An empirical expression for this function is given by \cite[e.g.][]{1991Natur.353..807H}
\begin{equation}
\label{fragmentation}
\frac{dN_{\pi}}{dz}=\frac{15}{16}z^{-3/2}(1-z)^2
\end{equation}
where $z=E/Q$. Detailed simulations have shown that the $\gamma$-ray spectrum from evaporating BHs is quite broad, peaking at $\sim 100$~MeV and cutting off at $E\approx m_{\pi}$ 
\citep[][]{1990PhRvD..41.3052M, 1991Natur.353..807H}.

If the lifetime of the BH is large compared with the observation time then the instantaneous flux of $\gamma$-ray photons (primary and secondary) emitted by a BH above some energy threshold $E_{D}$ can be written as \citep[][]{1991Natur.353..807H}
\begin{eqnarray}
\label{halzen17}
\frac{dN_{\gamma}}{dt}(>E_{D})\approx 8.0\times10^{23}\frac{Q}{1~\mathrm{GeV}}\nonumber
\\\times \left[\frac{1}{4}\left(\frac{Q}{E_{D}} \right)^{1/2}\left(1+\frac{E_{D}^{2}}{Q^{2}}\right)\right.\nonumber 
\\\left.-1+\frac{3}{2}\left(\frac{E_{D}}{Q}\right)^{1/2}-\frac{E_{D}}{Q} \right]s^{-1}
\end{eqnarray}
where $Q$ is the peak energy of the quark flux expressed in $GeV$.

\section{Detection of BH radiation}
\label{sec:The possibility of direct detection of BHs}

\subsection{Maximum distance for detection (general)}

The BH brightness, i.e., the energy emitted per unit time, per unit area, per unit frequency $\nu$, can be written, multiplying equation (\ref{particle_emission_photon}) by the photon energy $h\nu$ and dividing it by the BH surface area $4\pi r_{s}^{2}$ (where $r_{s}=2GM/c^2$) as
\begin{equation}
\label{equ-brigthness}
B_{\nu}(T) = \frac{\Gamma_{s} h\nu}{4\pi r_{s}^{2}}\left[\exp\left(\frac{8\pi^2 r_{s}\nu}{c}\right)-1\right]^{-1}.
\end{equation}

Let $S_{\nu}$ be the flux density (energy arriving per unit time, per unit frequency $\nu$) reaching a detector placed at some distance $d$ from the centre of the BH. The values of $B_{\nu}$, $S_{\nu}$ and $d$ are then related by \citep[e.g.][]{1999acfp.book.....L}
\begin{equation}
\label{equ-fluxdistance}
S_{\nu}=\Omega_{s}B_{\nu}(T)\approx\frac{\pi r_{s}^{2}}{d^{2}}B_{\nu}(T)
\end{equation}
where $r_{s}$\space is the Schwarzschild radius and $\Omega_{s}$ is the solid angle subtended by the BH in the sky. Considering $S_{\nu}$ equal to some detector sensitivity, then the distance $d$ is the \emph{maximum distance} at which we can detect a Schwarzschild BH of radius $r_{s}$ at a particular frequency $\nu$. We get from equation (\ref{equ-fluxdistance}) 
\begin{equation}
\label{equ-maxdistancenew}
d_{max}=\left(\frac{\pi r_{s}^{2}}{S_{\nu}}B_{\nu}(T)\right)^{1/2}
\end{equation}
In practice, a detector operates at some frequency band $[\nu_{1},\nu_{2}]$ which means that it is more accurate to consider 
\begin{equation}
\label{equ-maxdistancenew-integral}
d_{max}=\left(\frac{\pi r_{s}^{2}}{S_{\nu}}\frac{1}{\Delta \nu}\int_{\nu_{1}}^{\nu_{2}}B_{\nu}(T)d\nu\right)^{1/2}
\end{equation}
where $\Delta \nu =\nu_{2}-\nu_{1}$. When the emission stays within the some order of magnitude along the entire bandwith it is sufficient to consider the value of $B_{\nu}(T)$ evaluated at the central frequency (equation \ref{equ-maxdistancenew}). When the emission peak is far from this central frequency it is more accurate to use equation (\ref{equ-maxdistancenew-integral}) instead.

\subsection{Maximum distance for detection ($\gamma$-rays)}

The present day $\gamma$-ray detectors count individual photons over the background instead of measuring energy fluxes. The detectable number of $\gamma$-ray photons per unit time, emitted by a BH at some distance $d$ and zenith angle $\theta$ is given by \cite[e.g.][]{1993PhRvL..71.2524A, 2010JETPV3P406}
\begin{equation}
\label{photons-gamma-ray-detector}
n_{\gamma}(d,\theta,t)=\frac{1}{4\pi d^{2}}\int_{E_{1}}^{E_{2}}\frac{d^{2}N_{\gamma}}{dEdt}A(E,\theta)dE
\end{equation}
where $\frac{d^{2}N_{\gamma}}{dEdt}$ is the number of photons emitted on $[E,E+dE]$ per unit time, $A(E,\theta)$ the effective area of the detector as a function of the $\gamma$-ray photons energy and the zenith angle $\theta$, and $[E_{1},E_{2}]$ represents the detector energy bandwidth.

For simplicity, we assume that the source is always near the zenith ($\theta\approx 0^{\circ}$) and that the effective area at normal incidence remains constant for the entire bandwidth. We can then write equation (\ref{photons-gamma-ray-detector}) in the form 
\begin{equation}
\label{photons-gamma-ray-detector2}
\frac{n_{\gamma}}{A}\approx \frac{1}{4\pi d^2}\left(\frac{dN_{\gamma}}{dt}(>E_{1})-\frac{dN_{\gamma}}{dt}(>E_{2})\right).
\end{equation}
The left hand side of equation (\ref{photons-gamma-ray-detector2}) gives us the number of $\gamma$-ray photons reaching the detector per unit time per unit area. This value decreases as one moves away from the BH. Eventually there is a particular distance for which the left hand side of equation (\ref{photons-gamma-ray-detector2}) equals the detector sensitivity (which, in the case of $\gamma$-ray detectors, is usually expressed in 
$\mathrm{~ph~cm}^{-2}\mathrm{s}^{-1}$). This distance corresponds to the maximum distance at which we can detect the BH with some particular detector. Thus, replacing the entire left hand side of equation (\ref{photons-gamma-ray-detector2}) by that detector sensitivity $S$ we get for the maximum distance of detection 
\begin{equation}
\label{equ-maxdistance_gamma}
d_{max}=\left[\frac{1}{4\pi S}\left(\frac{dN_{\gamma}}{dt}(>E_{1})-\frac{dN_{\gamma}}{dt}(>E_{2})\right)\right]^{(1/2)}.
\end{equation}

\subsection{Current detector sensitivities from radio to $\gamma$-ray}

If the sensitivity comes in magnitudes (as it is usual at UV, optical, and IR wavelengths; through the use of a filter -- e.g. Johnson $B$ \citep[][]{1966ARA&A...4..193J}) we must convert it to a flux density. In order to express an apparent magnitude $m_{a}$ as a flux density $S_f$ we use the expression \citep[e.g.][]{zom90}
\begin{equation}
\label{equ_magflux}
m_{a}-m_{0}=-2.5 \log\frac{S_f}{S_{0}}
\end{equation}
where $m_{0}$ is a reference magnitude (usually $m_{0}=0$ is defined for the star Vega at all colours/filters) and $S_{0}$ is the corresponding flux density. If $m_{a}$ corresponds to the limiting magnitude of the telescope for a filter $X$ with central wavelength $\lambda_X$ and bandwidth $2\Delta\lambda_X$ (defined at half the peak through-flux), then $S_f$ will be the corresponding sensitivity $S_{\nu}$, at $\nu_X=c/\lambda_X$ (with propagating error $\Delta \nu_X = c \Delta\lambda_X / \lambda_X^2$).

In what follows we particularize our study by separating the electromagnetic spectrum into six bands, using the currently known limitations in sensitivity for each one (Table~\ref{tbl-1}).

\begin{table*}
\begin{minipage}{140mm}
\caption{The best sensitivities available for a range of wavelengths inside the six major bands of the electromagnetic spectrum (column {\bf (1)}): {\bf (2):} central wavelength (the Johnson filter is indicated for the optical); {\bf (3):} central frequency ($\nu_X=c/\lambda_X$); {\bf (4):} sensitivity, obtained from the references of column (7), as explained in the text; {\bf (5):} bandwidth; {\bf (6):}  the most sensitive telescope, for each observing wavelength; {\bf (7):} references from where the values were taken:
[1] \citet[][]{2006astro.ph.10596R};
[2] \citet[][]{2011ApJ...739L...1P};
[3] \citet[][]{2010A&A...518L...3G};
[4] \citet[][]{2010A&A...518L...2P};
[5] \citet[][]{2004ApJS..154...10F};
[6] \citet[][]{Ubeda2012,Dressel2012};
[7] \citet[][]{2007ApJS..173..682M};
[8] \citet[][]{2001A&A...365L..45H};
[9] \citet[][]{2003A&A...411L.131U};
[10] \citet[][]{2009ApJ...697.1071A};
[11] \citet[][]{2010APh....34..245R}.} 
\label{tbl-1}
 \centering
\begin{tabular}{rcccccc}
\hline
\\
{\bf (1)} & {\bf (2)} & {\bf (3)} & {\bf (4)} & {\bf (5)} & {\bf (6)} & {\bf (7)}  \\
Band & $\lambda_{X}$  &  $\nu_{X}$ (Hz) & $S_{\nu}$ & Bandwidth & Telescope & Ref.\\
\\
\hline
\\
Radio & 20~m & $1.5\times10^7$ &$1.1\times10^{-2}$~Jy & 4~MHz  & LOFAR & [1] \\
& 1.5~m & $2.0\times10^8$ &$6.3\times10^{-5}$~Jy & 4~MHz  & LOFAR & [1] \\
& 3.5~cm & $9.0\times10^9$ & $1.0\times10^{-6}$~Jy & 3~MHz  & e-VLA & [2] \\

Infrared & 500~$\mu$m & $6.0\times10^{11}$ &$6.8\times10^{-3}$~Jy & 368~$\mathrm{\mu m}$&  Herschel & [3]  \\
& 70~$\mathrm{\mu m}$ & $4.3\times10^{12}$ &$4.4\times10^{-3}$~Jy & 25~$\mathrm{\mu m}$ &  Herschel & [4] \\
& 3.58~$\mathrm{\mu m}$ & $8.4\times10^{13}$ &$4\times10^{-7}$~Jy & 0.75~$\mathrm{\mu m}$  & SST & [5] \\

Visible & 0.55~$\mathrm{\mu m}$ (V) & $5.5\times10^{14}$ & $3.7\times10^{-9}$~Jy & 0.089~$\mu$m   & HST & [6] \\

UV & 232~nm & $1.3\times10^{15}$ &$1.8\times10^{-8}$~Jy & 106~nm  &GALEX& [7] \\
& 154~nm & $1.9\times10^{15}$ &$1.6\times10^{-9}$~Jy & 44~nm  &GALEX & [7] \\

X-rays & 3.5~nm &$8.6\times10^{16}$ &$5.5\times10^{-10}$~Jy & 0.3~keV  & XMM & [8] \\

& 1~nm & $3.0\times10^{17}$&$8.5\times10^{-11}$~Jy & 1.5~keV &  XMM & [8] \\

& 0.165~nm & $1.8\times10^{18}$ & $2.0\times10^{-10}$~Jy & 5~keV  & XMM &  [8] \\

$\gamma$-rays & $2.5\times10^{-13}$~m & $1.2\times10^{21}$ & $5.0\times10^{-7}\mathrm{~ph~cm}^{-2}\mathrm{s}^{-1}\mathrm{keV}^{-1}$ & 9.8~MeV  & INTEGRAL & [9]  \\

& $8.3\times10^{-18}$~m & $3.6\times10^{25}$ & $3.0\times10^{-9}\mathrm{~ph~cm}^{-2}\mathrm{s}^{-1}$ & 300~GeV  & Fermi & [10] \\

& $2.5\times10^{-19}$~m & $1.2\times10^{27}$ & $5.0\times10^{-11}\mathrm{~ph~cm}^{-2}\mathrm{s}^{-1}$ & 9.9~TeV  & HESS & [11]\\
\hline
\end{tabular}
\end{minipage}
\end{table*}

There are, at present, several radio telescopes and interferometer arrays operating from the millimeter and sub-mm to metric waves. Sensitivities vary according to the detector characteristics  and, for each one, with the observing wavelength. On the metric wave domain the most sensitive radio telescope is the Low Frequency Array (LOFAR). It operates from $\lambda = 1.2\mathrm{~m}$ (240~MHz) up to $\lambda=20\mathrm{~m}$ (15~MHz) with sensitivities that vary from $\approx 0.07$~mJy ($1.2\mathrm{~m}\leq \lambda\leq 2.5\mathrm{~m}$) to 11~mJy at $\lambda = 20\mathrm{~m}$ \citep[][]{2006astro.ph.10596R}. We, then, picked the wavelengths 20~m (the longest one) and 1.5~m (the most sensitive).

Moving to centimetric wavelengths we have as the most sensitive radio telescope the expanded Very Large Array (e-VLA) which provides a complete coverage from 1~GHz to 50~GHz ($0.6\mathrm{~cm}\leq \lambda\leq 30\mathrm{~cm}$) with a continuum sensitivity of, typically, $1~\mu\mathrm{Jy}$ which is $\approx 10$ times more sensitive than its predecessor, the Very Large Array \citep[][]{2011ApJ...739L...1P}. We consider the detection at $\lambda=3.5\mathrm{~cm}$ ($\approx 9\mathrm{~GHz}$).

The Herschel Space Observatory
\footnote{Although already defunct, data mining from this observatory will carry on for many years.} covers the far-infrared band of the spectrum reaching submillimetric wavelenghts \citep[$60\mathrm{~\mu m}\leq \lambda\leq 671\mathrm{~\mu m}$ -- ][]{2010A&A...518L...3G,2010A&A...518L...2P}. We picked two wavelengths in the above range: 500~$\mu$m (submillimeter) and 70~$\mu$m (far-infrared) and considered as maximum sensitivities, respectively, 6.8~mJy \citep[][]{2010A&A...518L...3G} and 4.4~mJy \citep[][]{2010A&A...518L...2P}. The Spitzer Space Telescope (SST), covers the near-infrared band (now that it is working in its warm phase). We picked $3.58\mathrm{~\mu m}$ (the shortest wavelength operated by SST) and considered a maximum sensitivity of $0.4\mathrm{~\mu Jy}$ \citep[][]{2004ApJS..154...10F}.

The sensitivity of optical detectors is normally expressed in terms of apparent magnitude and the most sensitive reaches $\sim 30$~mag \citep[the Advanced Camera for Surveys and Wide Field and Planetary Camera 3 on the Hubble Space Telescope;][]{Ubeda2012,Dressel2012} for all of the equivalent Johnson filters BVR \citep[][]{1966ARA&A...4..193J}: 0.44~$\mu$m, 0.55~$\mu$m, and 0.70~$\mu$m, respectively. Refering to Vega, $m_0=0$ corresponds, in the case of the filter V, to $S_0=3670$~Jy \citep[]{zom90}. Then, from equation~(\ref{equ_magflux}), with $m_a=30$~mag, we get $S_\nu=3.7\times10^{-9}$~Jy. For filters B and R we get similar results. We, then, picked filter V which is central to the optical band.

The instruments onboard the Galaxy Evolution Explorer
(GALEX),  were able to detect radiation at 232~nm (near-UV) and 154~nm (far-UV) down to a limiting magnitude of, respectively, $m_{a}=24.4$ and $m_{a}=24.8$. The white dwarf LDS749B is used as the primary GALEX standard with the reference magnitudes $m_{0}=20.1$ (232~nm) and $m_{0}=18.8$ (154~nm) \citep[][]{2007ApJS..173..682M}. The radiation flux of LDS749B is $2.6835\times10^{-14}\mathrm{~ergs}^{-1}\mathrm{cm}^{-2}$ \AA$^{-1}$ when $\lambda\approx 232\mathrm{~nm}$ and $2.6676\times10^{-14}\mathrm{~ergs}^{-1}\mathrm{cm}^{-2}$ \AA$^{-1}$ when $\lambda\approx 154\mathrm{~nm}$ \citep[][]{2008AJ....135.1092B} which corresponds to, respectively, $S_{0}\approx 9.5\times10^{-7}\mathrm{~Jy}$ and $S_{0}\approx 3.9\times10^{-7}\mathrm{~Jy}$ (multiplying the given values by the corresponding bandwidths -- see Table~\ref{tbl-1} -- and converting the obtained results to Wm$^{-2}$s). Making use of equation (\ref{equ_magflux}) we get the sensitivities $S_{\nu}=1.8\times10^{-8}\mathrm{~Jy}$ (232~nm) and $S_{\nu}=1.6\times10^{-9}\mathrm{~Jy}$ (154~nm).
 
The most sensitive X-ray telescope operating in orbit is the X-ray Multi-mirror Mission (XMM-Newton). It has, for example, a sensitivity of $S_{\nu}=5.5\times10^{-10}$~Jy at $0.2$--$0.5$~keV \citep[soft X-rays;][]{2001A&A...365L..45H}, 
$S_{\nu}=8.5\times10^{-11}$~Jy  
at $0.5$--$2.0$~keV (soft and mid X-rays), and  
$S_{\nu}=2.0\times10^{-10}$~Jy  
at $5$--$10$~keV (mid X-rays). Sensitivities in $Jy$ were obtained dividing the given value in $\mathrm{erg}\mathrm{s}^{-1}\mathrm{cm}^{-2}$ (after conversion to Wm$^{-2}$) by the corresponding bandwidth. For example, at  $0.5$--$2.0$~keV we have \citep[see][]{2001A&A...365L..45H} $S_{\nu}=3.1\times10^{-16}\mathrm{~erg}\mathrm{s}^{-1}\mathrm{cm}^{-2}=3.1\times10^{-19}\mathrm{~Wm}^{-2}$. Dividing this value by the bandwidth $\Delta \nu \approx 3.6\times10^{17}\mathrm{~Hz}$ (which corresponds, in terms of energy, to 1.5~keV) we get 
$S_{\nu}=8.5\times10^{-37}\mathrm{~Wm}^{-2}\mathrm{s}=8.5\times10^{-11}\mathrm{~Jy}$. 
We, then, picked the wavelengths  3.5~nm (soft X-rays), 1~nm (soft and mid X-rays), and 0.165~nm (mid X-rays) which are central to the considered bands -- see Table~\ref{tbl-1}. 

In the soft $\gamma$-ray domain the detectors on board the INTErnational Gamma-Ray Astrophysics Laboratory (INTEGRAL) allow for observations between 175~keV (hard X-rays) and 10~MeV (soft $\gamma$-rays) with a typical sensitivity of  $\approx 5.0\times10^{-7}\mathrm{~ph~cm}^{-2}\mathrm{s}^{-1}\mathrm{keV}^{-1}$ \citep[][]{2003A&A...411L.131U}. In the medium $\gamma$-ray domain the Fermi Gamma-ray Space Telescope allows for observations between 20~MeV and 300~GeV with a sensitivity of  $\approx 3.0\times10^{-9}\mathrm{~ph~cm}^{-2}\mathrm{s}^{-1}$ \citep[][]{2009ApJ...697.1071A}. 
Operating at hard $\gamma$-rays, the High Energy Stereoscopic System (HESS) has a sensitivity that varies from $\approx 5.0\times10^{-11}\mathrm{~ph~cm}^{-2}\mathrm{s}^{-1}$ at $0.1$~TeV down to $\approx 5.0\times10^{-14}\mathrm{~ph~cm}^{-2}\mathrm{s}^{-1}$ at $10$~TeV \citep[][]{2010APh....34..245R}. Since we are interested in PBHs with more than $10^{-22}$~M$_{\sun}$ (see discussion in Section \ref{sec:Results}), i.e., PBHs with the emission peak $Q < 0.2$~TeV (cf. equation \ref{equ-temp2}), we consider for the HESS sensitivity the value $5.0\times10^{-11}\mathrm{~ph~cm}^{-2}\mathrm{s}^{-1}$. 
We, then, picked the wavelengths  $2.5\times10^{-13}$~m (soft $\gamma$-rays), $8.3\times10^{-18}$~m (mid $\gamma$-rays), and $2.5\times10^{-19}$~m (strong $\gamma$-rays) which are central to the considered bands -- 
see Table~\ref{tbl-1}.

\section{Results (Maximum Distance for Detection)}
\label{sec:Results}

\begin{figure}
\centering
\includegraphics[width=84mm]{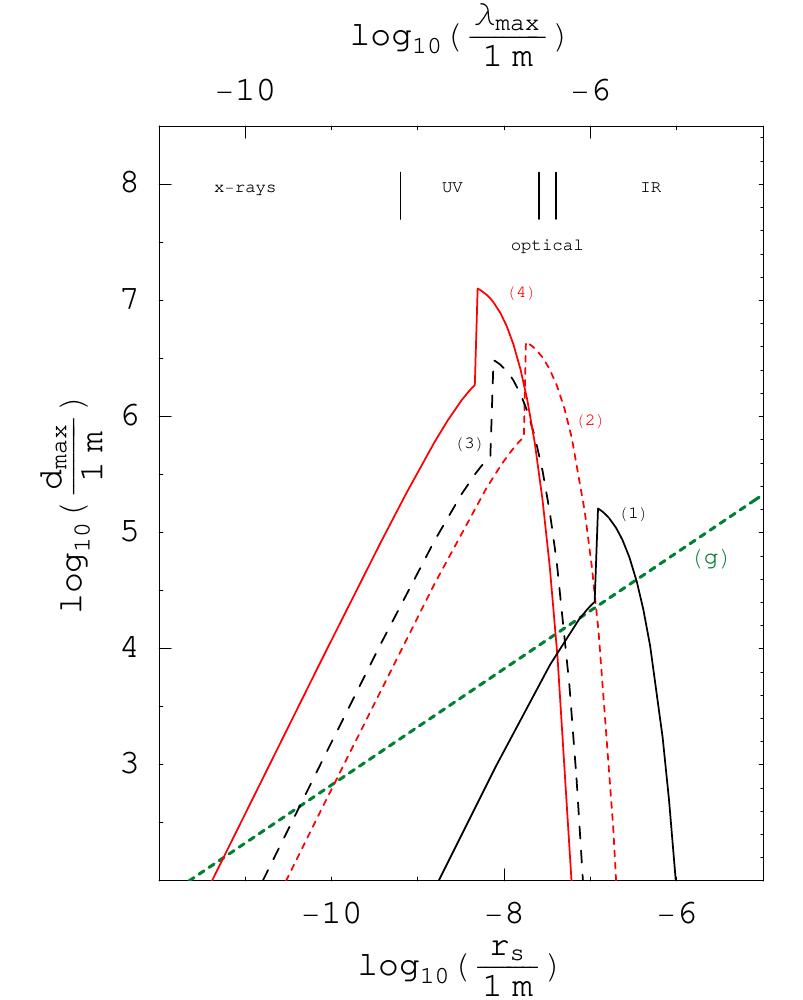}
\caption{{\bf[IR - Optical - UV]} Maximum distance (d$_{max}$) for detecting the electromagnetic radiation directly emitted by a BH as a function of the Schwarzschild radius ($r_{s}$) for the wavelengths:
(1) 3.58~$\mu$m
(2) 0.55~$\mu$m;
(3) 232~nm;
(4) 154~nm
(cf. Table \ref{tbl-1}). At the top, the horizontal axis is also divided in terms of BH colours ($\lambda_{max}$; equation (\ref{equ-wien})). For reference, the dashed line labeled (g) represents the distance for which the detector would be subject to a gravitational acceleration of 1~g. See the text for more details.}
\label{figure1}
\end{figure}

\begin{figure}
\centering
\includegraphics[width=84mm]{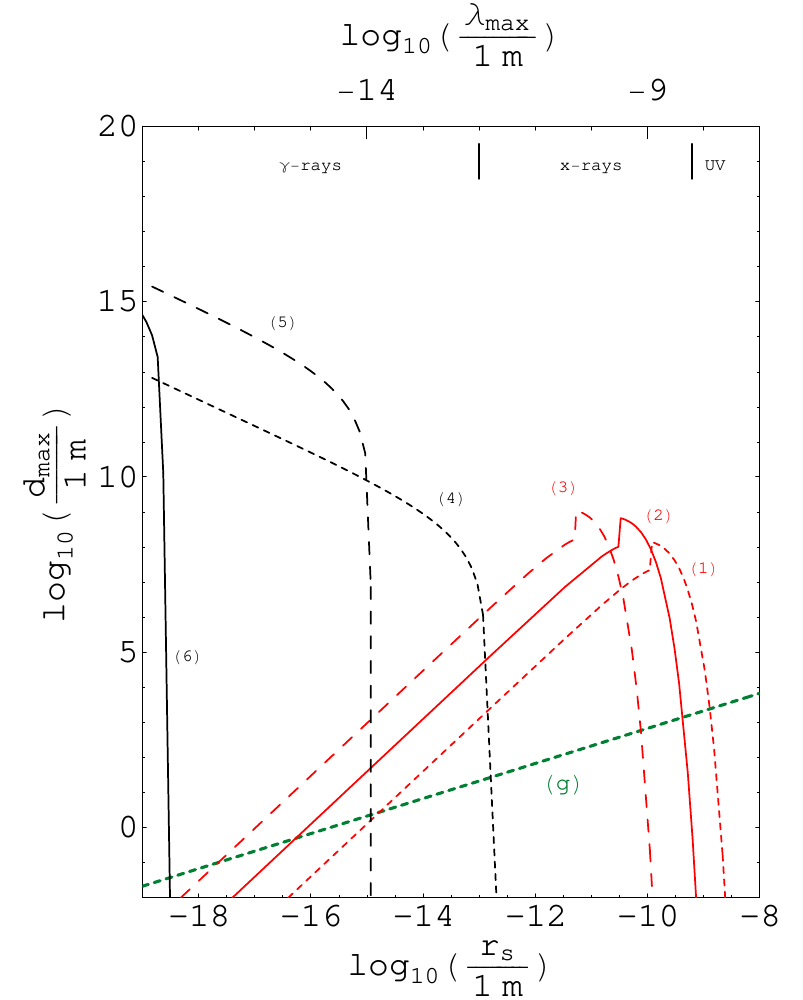}
\caption{{\bf[X-rays - $\gamma$-rays]} Maximum distance (d$_{max}$) for detecting the electromagnetic radiation directly emitted by a BH as a function of the Schwarzschild radius ($r_{s}$) for the wavelengths:
(1) 3.5~nm;
(2) 1~nm;
(3) 0.165~nm;
(4) $2.5\times10^{-13}$~m;
(5) $8.3\times10^{-18}$~m;
(6) $2.5\times10^{-19}$~m
(cf. Table \ref{tbl-1}).
In cases (4), (5) and (6) we do not see a peak since the main contribution for the total emitted flux is from secondary $\gamma$-rays.
See the text and caption of Figure \ref{figure1} for more details.}
\label{figure2}
\end{figure}

In Figures \ref{figure1} and \ref{figure2} we present the maximum distance $d(r_{s})$ at which some detector working with sensitivity $S_{\nu}$ (within our current technologies, cf. Table \ref{tbl-1})  at a particular wavelength $\lambda$ should be placed in order to detect the emission of a given BH (from equations (\ref{equ-maxdistancenew}), (\ref{equ-maxdistancenew-integral}), and (\ref{equ-maxdistance_gamma})).

For the longest wavelengths we found out that the detector should be quite near the BH in order to detect the corresponding emission which means that it would be under the influence of a very strong gravitational field. In fact, for the first five cases on Table~\ref{tbl-1} the detection would not be pratical. For example, when $\lambda = 70~\mu \mathrm{m}$ the detector should be placed at a distance of $\approx 400\mathrm{~m}$ in order to detect the IR emission of a $\sim 10^{21}\mathrm{~kg}$ BH, in which case it would be under a gravity acceleration of $\sim 10^{6}\mathrm{~ms}^{-2}$. When $\lambda = 3.58~\mu \mathrm{m}$ the detector should be placed at a distance of $\approx 170\mathrm{~km}$ in order to detect the IR emission of a $\sim 10^{20}\mathrm{~kg}$ BH, which is now a safe distance (the detector would experience a gravity acceleration of only $\sim 0.18\mathrm{~ms}^{-2}$). For the remaining cases the gravitational effects are much smaller ($<10^{-4}\mathrm{~ms}^{-2}$) and the detectors would all be located at safe distances.

Since smaller BHs evaporate in less than $\sim 1\mathrm{~month}$ (equation (\ref{equ-evaptime})) giving rise to a $\gamma$-ray burst, we thus considered the detection of BHs with Schwarzschild radius from $\sim 10^{-19}\mathrm{~m}$ ($\sim 10^{-22}$~M$_{\sun}$) to $\sim 10^{-7}\mathrm{~m}$ ($\sim 10^{-11}$~M$_{\sun}$) -- present day values.
It is not our objective to study the final stages of BH evaporation, for which published results abound (see Section \ref{sec:Introduction}).

From the near infrared to X-rays, the peak in each curve (Figures \ref{figure1} and \ref{figure2}) corresponds to the BH that can be detected farthest for the corresponding ($\lambda$, $S_{\nu}$) pair. It is possible to detect BHs farther and farther as we move our observations into shorter and shorter wavelengths. On Table \ref{tbl-max} we show, for each of these BHs, the values of ($r_{s}$, $d_{max}$) (equations (\ref{equ-maxdistancenew}) or (\ref{equ-maxdistancenew-integral})) as well as the corresponding BH mass ($r_{s}=2GM/c^2$) and temperature (equation (\ref{equ-temp2})). As regards $\gamma$-rays (Figure \ref{figure2}, curves (4), (5), and (6)) it is possible to detect BHs with masses ranging from 
 $\sim 10^{14}\mathrm{~kg}$ ($\sim 10^{-16}$~M$_{\sun}$) at distances of $\sim 10^{8}\mathrm{~m}$ down to $\sim 10^{8}\mathrm{~kg}$ ($\sim 10^{-22}$~M$_{\sun}$) at distances of $\sim 3\times10^{15}\mathrm{~m}$ ($\approx 0.1\mathrm{~pc}$).

In summary, we get the following result: 
$[10^{-22},10^{-11}]$M$_{\sun}$ BHs (present day masses)  can be detected at 
$d_{max}~=~[0.1\mathrm{~pc},~10^{5}\mathrm{~m}]$.

\begin{table*}
\begin{minipage}{120mm}
\caption{The maximum distances for direct detection of BHs for the wavelengths considered on Table \ref{tbl-1} (column {\bf (2)}), only for the cases when the detector cannot be significantly influenced by the BH gravity. For each wavelength we show the BH that can be detected the farthest (corresponding to the peak of the respective curve on Figures \ref{figure1} or \ref{figure2}): {\bf (2)}: Schwarzschild radius; {\bf (3)} and {\bf (4)}: mass; {\bf (5)}: temperature; {\bf (6)}: maximum distance for detection; {\bf (7)}: figure where each curve appears. See further explanations on the main text.} 
\label{tbl-max}
\centering
\begin{tabular}{ccccccc}
\hline
\hline
\\
{\bf (1)} & {\bf (2)} & {\bf (3)} & {\bf (4)} & {\bf (5)} & {\bf (6)} & {\bf (7)} \\
$\lambda_{X}$  &  $r_{s}$~(m) & $M$~(kg)& $M$~($M_{\sun}$) & T~(K) & $d_{max}$~(m) & Figure\\
 \\
\hline
\\

3.58~$\mathrm{\mu m}$ & $1.1\times10^{-7}$ & $7.7\times10^{19}$ & $3.9\times10^{-11}$ & $1.6\times10^{3}$  & $1.7\times10^5$  & 1\\

0.55~$\mathrm{\mu m}$ (V) & $1.8\times10^{-8}$ & $1.2\times10^{19}$ & $5.9\times10^{-12}$ & $1.0\times10^{4}$   & $4.4\times10^6$ & 1\\

232~nm & $7.4\times10^{-9}$ &$5.0\times10^{18}$ & $2.5\times10^{-12}$ & $2.5\times10^{4}$  & $3.1\times10^6$ & 1 \\

154~nm & $5.8\times10^{-9}$ &$3.9\times10^{18}$ & $1.9\times10^{-12}$ & $3.2\times10^{4}$  & $1.2\times10^7$ & 1\\

3.5~nm &$1.1\times10^{-10}$ &$7.6\times10^{16}$ & $3.8\times10^{-14}$ & $1.6\times10^{6}$  & $1.4\times10^8$ & 2 \\

1~nm & $3.3\times10^{-11}$& $2.2\times10^{16}$ & $1.1\times10^{-14}$ & $5.5\times10^{6}$ &  $6.7\times10^8$ & 2 \\

0.165~nm & $5.3\times10^{-12}$ & $3.6\times10^{15}$ & $1.8\times10^{-15}$ & $3.4\times10^{7}$ & $1.1\times10^9$  & 2\\

\hline
\end{tabular}
\end{minipage}
\end{table*}

\section{Discussion}

BHs are objects predicted by the Laws of Physics. In terms of mass they could have from a few Planck masses ($M_P\sim 10^{-8}\mathrm{~kg}$) up to $\sim 10^{10}$~M$_{\sun}$. SMBHs have already been identified in the centre of many galaxies, including our own. In terms of stellar mass, strong candidates have been identified in binary systems. All, however were detected via indirect means (e.g. dynamical interactions with their surroundings).

Sub-stellar mass BHs might have been produced during the early stages of the Universe. In particular, those with an initial mass of $\sim 10^{12}\mathrm{~kg}$ or greater should still be lurking around us at the present time (smaller ones should have already completely evaporated). There are hopes of directly detecting these PBHs via the electromagnetic radiation that they emit.

It was this path that we explored in this paper. In summary, such a BH must: a) have a mass on $[10^{-22},10^{-11}]$~M$_{\sun}$, i.e., it must be primordial and not have already exploded \citep[e.g.][]{2010PhRvD..81j4019C}; 
 b) radiate according to equation (\ref{particle_emission});
 c) be a Schwarzschild BH (not being one will reduce its chances of detection). 

We have assumed present detector technologies, as regards sensitivity (conservative, since these will become better in the future). It was our aim to show that, using detectors based on these technologies, it might be possible to look for signatures of individual BHs (not necessarily at the end stage of their evaporation process when they explode giving rise to a $\gamma$-ray burst). 
 We point out three paradigmatic examples:

\begin{itemize}
\item A BH like the fourth in Table~\ref{tbl-max} ($\sim 10^{18}\mathrm{~kg}$; 10$^{-12}$~M$_{\sun}$) is detectable in the optical {\em and} UV up to $\sim 10^{7}\mathrm{~m}$: in detail, these detection possibilities are seen in curves (2,3,4) of Figure~\ref{figure1} ($r_{s}\sim 10^{-8}\mathrm{~m}$).
The evaporation time of such a BH is $\sim 10^{29}$~yr (cf. equation \ref{equ-evaptime}) which means that it would emit a steady and continuous flux for a long time at all wavelengths. Good sampling of the emission at IR, optical and UV wavelengths could be enough to unequivocally identify the source of emission as a PBH, since the shape of the emission curve as given by equation (\ref{particle_emission})  is quite unique.
\item  A BH with $\sim 10^{11}\mathrm{~kg}$ (10$^{-19}$~M$_{\sun}$; $r_{s}\sim 10^{-16}\mathrm{~m}$; $t_{evap}\sim 10^{15}\mathrm{~s}$) can be detected in $\gamma$-rays at a maximum distance of $\sim 10^{11}\mathrm{~m}$ ($\approx 0.7$~au) or $\sim 10^{13} \mathrm{~m}$ ($\approx 70$~au) depending on the detector used (Figure~\ref{figure2}, the points at the extreme left of the curves (4,5)). Such a detection will open the path to study the emission curve within the $\gamma$-ray domain. 
\item  A BH with $\sim 10^{8}\mathrm{~kg}$ (10$^{-22}$~M$_{\sun}$; $r_{s}\sim 10^{-19}\mathrm{~m}$) can be detected in $\gamma$-rays at a maximum distance of 0.03~pc or 0.1~pc depending on the detector used (Figure~\ref{figure2}, curves (5,6)). Such a BH would explode in a few decades (or less) which means that monitoring its flux levels could identify it as a PBH and even allow a prediction of the time of its explosion.
\end{itemize}

Laboratorial detection of electromagnetic radiation emitted by Planckian-size BHs might also be possible. In fact, if we live in a Universe with more than three spatial dimensions then it is expected that these Planckian-size BHs might be produced at the LHC. The detection of a BH at the LHC would provide a first experimental and secure test of the electromagnetic radiation emission mechanism. 

Focusing now on practicalities, the farther we look into the Universe (the larger the volume sampled) the better our chances of finding a small-mass BH. It seems, then, that the best waveband to start searching for BHs is $\gamma$-rays $d_{max}\leq~0.1$~pc). However we do recomend that the detection should be attempted on the other electromagnetic spectrum bands as well. On the X-ray band, for example, we migth be able to detect BHs at distances of the order of the Earth-Moon distance.

In \citet[][]{sobrinho2011} it was shown that, within some scenarios, it is expected a number of $\sim 10^5$ PBHs within the Oort cloud. For these PBHs, mainly with $\sim 10^{18}$~g, the electromagnetic emission peaks at the hard X-ray band. It is plausible that, throughout the history of the Solar System, some PBHs might have been thrown into the inner Solar System following the same fate as the comets (i.e., we could have PBHs describing elliptical orbits with semi-major axis $100$--$200$~au). That would improve the chances of direct detection of a PBH by a space probe. For example, NASA is planing to launch, possibly in 2014, the \emph{Innovative Interstellar Explorer} (IIE) a mission that is expected to reach 200~au in about 30 years after launch \citep[][]{2006AIPC..858..341M}. 

We will need to study the cosmological density of small-mass BHs in order to find the probability of detecting these directly, given the results of this paper (Sobrinho \& Augusto, in prep.). 
Then, we will have a more complete idea on the potential of current and future space missions as regards direct detection of BHs.

\section{Acknowledgements}

The authors are very grateful to Jos\'{e} Sande Lemos for his encouragement on writing this paper.  His comments, as a member of the jury during the PhD defense of the first author, were very stimulating. 
 We also acknowledge helpful discussions with Alastair Edge and an anonymous referee whose valuable comments improved this paper substantially. Finally, we acknowledge the clarifications given by Jane MacGibbon, Francis Halzen and Enrique Zas.

\end{document}